\documentclass[preprint,showkeys,preprintnumbers,amsmath,amssymb]{revtex4-1}

\usepackage{graphicx}
\usepackage{dcolumn}
\usepackage{bm}

\begin{document}

\title{Precise Optical Measurement of Carrier Mobilities \\ Using \emph{Z}-scanning Laser Photoreflectance}

\author{Will Chism}
 \email{e-mail address: wchism@xitronixcorp.com}
\affiliation{%
Xitronix Corporation\\
106 E. Sixth St., Ninth Floor, Austin, TX, USA 78701
}%

\date{\today}

\begin{abstract}
A simple yet precise optical technique for measuring the ambipolar carrier mobility in semiconductors is presented.  Using tightly focused Gaussian laser beams in a photo-reflectance system, the modulated reflectance signal is measured as a function of the \emph{Z} (longitudinal) displacement of the sample from focus.  The modulated component of the reflected probe beam is a Gaussian beam with its profile determined by the focal parameters and the complex diffusion length. The reflected probe beam is collected and input to the detector, thereby integrating over the radial profile of the beam.  This results in analytic expressions for the \emph{Z} dependence of the signal in terms of diffusion length and recombination lifetime.  Best fit values for the diffusion length and recombination lifetime are obtained via an iterative fitting procedure. The output diffusion lengths and recombination lifetimes and their estimated uncertainties are combined according to the Einstein relation to yield the mobility and its uncertainty.

\end{abstract}


\maketitle

\section{\label{sec:level1}Introduction}

Carrier mobility is a key electronic property influencing semiconductor device performance. In semiconductor device manufacturing, many process have the potential to alter or degrade carrier mobility.  Thus the ability to measure the carrier mobility between processing steps, in a nondestructive manner, is advantageous in the semiconductor industry. Numerous optical techniques have been developed for measurement of carrier electronic properties. Surface photovoltage (SPV) \cite{SPV,SPV2}, photoconductance decay \cite{FCAetc}, free carrier absorption \cite{FCAetc},  photoluminescence \cite{PL}, and time-resolved THz spectroscopy \cite{THz,Heilweil} are among the techniques frequently reported. The last technique, THz spectroscopy, is potentially sensitive to carrier mobility but requires a complex experimental setup.

In this letter, a simple yet precise optical technique for measuring carrier mobility in semiconductors is demonstrated. The technique is based upon profiling of the output signals of a laser photo-reflectance (LPR) system as the sample is stepped through focus.  The technique may be used to simultaneously characterize carrier diffusion length and recombination lifetime.  These carrier properties determine the diffusion coefficient, or equivalently, the carrier mobility.

To demonstrate this technique, referred to as \emph{Z}-scanning laser photo-reflectance, parameterized expressions for the LPR signal amplitude and phase were fit to experimental \emph{Z}-scan LPR data obtained from samples consisting of shallow electrical junctions formed in silicon.  Independent estimates of the diffusion length and recombination time were obtained from the fit procedure.  Statistical estimates of fit error were used to estimate precision of the determined diffusion lengths and recombination lifetimes.  These values were used to determine the mobility and its precision via the Einstein relation. Systematic effects of process variations on carrier electronic properties are observed. For the selected set of \emph{Z}-scanning measurement parameters, statistical uncertainties in the determined optical mobility are demonstrated at less than $2\%$.

\section{The \emph{Z}-scanning Laser Photo-Reflectance Technique}

Photo-reflectance refers to the use of an intensity modulated pump light beam to photo-inject charge carriers in a semiconductor sample while a second probe light beam is used to detect the modulated reflectance of the sample. Phase-locked detection of the signal at the known modulation frequency is used to suppress unwanted noise, resulting in the ability to detect reflectance changes at the ppm level \cite{Pollak}. The technique reported here uses Gaussian laser beams for both the pump and probe beams in a photo-reflectance apparatus.  The beams are collinear and co-focused to a point along the beam path. \emph{Z} is the displacement of the sample from the common beam waist along the focal column. The reflected probe beam is collected and input to the detector, thereby integrating over the radial profile of the beam. The remaining \emph{Z} dependence of the signal depends only upon the diffusion length, recombination time, and focal parameters.  The detector output is passed to the lock-in amplifier, which measures the LPR signal. The acquired LPR signal is the relative change in the (radially integrated) reflected probe light intensity and consists of a vector characterized by an amplitude and a phase. The amplitude is the induced (AC) change in reflectance divided by the DC (\emph{i.e.} linear) reflectance, whereas the phase characterizes the temporal delay of the reflectance change due to the relaxation dynamics of the carriers within the sample. Thus the LPR signal is acquired as a function of \emph{Z}.

\begin{figure}
\includegraphics[width=246pt]{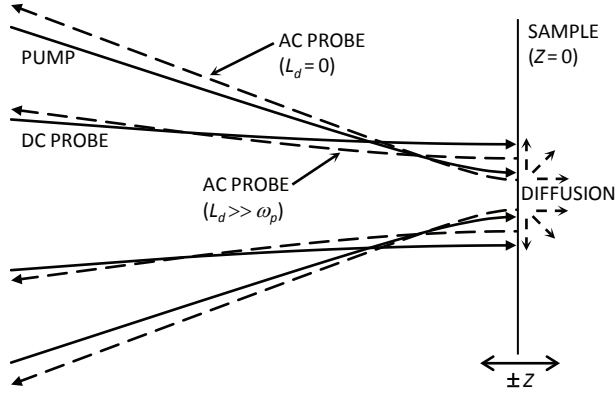}
\caption{\label{fig:focus}Representative cross-section of beams present in the Z-scanning LPR system.  The reflected AC probe beam profile is highly sensitive to $L_{d}$.}
\end{figure}

The focal geometry of the incident pump and probe beams at $Z=0$ is illustrated in Fig.~\ref{fig:focus}. At focus, the linear reflected pump and probe beam profiles will coincide with the respective input beams. The pump will induce a reflectance modulation within a radius $\omega_{m}\equiv (\omega_{p}^{2}+L_{d}^{2})^{1/2}$, where $\omega_{p}$ is the pump beam waist and $L_{d}$ is the diffusion length. The modulated component of the reflected probe beam is a Gaussian beam with radius defined by the incident probe beam radius and the radius of modulation.  As the beam is stepped through focus, the area of modulation varies according to the relation $\omega_{m}^{2}(Z)=\omega_{p}^{2}(Z)+L_{d}^{2}$, where $\omega_{p}(Z)=\omega_{p}\sqrt{1+(Z/z_{p})^{2}}$ is the radius of the incident pump beam as a function of \emph{Z} ($z_{p}=\pi\omega_{p}^{2}/\lambda_{p}$ is the Rayleigh range of the pump beam and $\lambda_{p}$ is the pump beam wavelength).   Thus a new reflected AC beam profile will be generated for each value of \emph{Z}.  Fig.~\ref{fig:focus} also shows the reflected AC probe beam profile for different values of $L_{d}$.  When $L_{d}\cong 0$, the waist of AC reflected probe beam is smaller than that of either the pump or the probe beam.  However, when $L_{d}^{2}\gg \omega^{2}$ the AC beam will approach the reflected DC profile. Integrating over the radial dependence of the reflected DC and AC beams results in simple analytic expressions for the \emph{Z} dependence of the LPR amplitude and phase in terms of carrier diffusion length and recombination lifetime.  These expressions may be used to determine $L_{d}$ and $\tau$ and their statistical uncertainties by nonlinear fitting to the \emph{Z}-scan LPR phase and amplitude data.  The mobility and its uncertainty can then be obtained from the Einstein relation $\mu = qD/k_{b}T$, where $D = L_{d}^{2}/\tau$ and $q/k_{b}T$ is the thermal voltage ($\cong26$ meV).

\section{Theoretical Principles}

At the critical points of a semiconductor material, the LPR signal arises from an electromodulation effect which exhibits a sharp third-derivative lineshape \cite{Aspnes72}. In this case the photo-reflectance signal becomes:
\begin{eqnarray}
\frac{\Delta R}{R}=\frac{2qN_{e}\Delta V}{\epsilon_{s}}\times L(\lambda)
\label{eq:one}.
\end{eqnarray}
where $q$ is the electronic charge, $N_{e}$ is the carrier concentration, $\Delta V$ is the SPV, $\epsilon_{s}$ is the static dielectric constant, and $L(\lambda)$ is a line-shape function determined by the semiconductor bandstructure ($\lambda$ is the probe beam wavelength).  Eq.~(\ref{eq:one}) is valid for depleted surfaces provided the electric field is not too inhomogeneous \cite{Aspnes72,Aspnes69}.

The SPV depends on the pump beam and the physics of its interaction with the sample \cite{SPV}.  The SPV is generally linear in the pump intensity provided the photo-injection is small with respect to the restoring current \cite{Schroder97,Park}. In this case the SPV will exhibit the spatial dependence of the excess carrier density. In the one-dimensional (1D) limit $\omega_{p}(Z)\gg L_{d}$ \cite{Opsal87}, the SPV may be obtained from the solution of the 1D differential equation for the modulated the carrier density \cite{SPV2}.  In the three-dimensional (3D) limit $\omega_{p}(Z)\ll L_{d}$ \cite{Opsal87}, the excess carrier density involves a zero-order Hankel transform of the 1D solution and therefore must be treated numerically \cite{Mandelis,Opsal83}. However, the use here of Gaussian laser beams for both the pump and probe means the reflected probe beam may instead be treated directly according to the analytically tractable method of Gaussian decomposition \cite{Weaire,SBahae,Petrov}.

Consider cylindrically symmetric Gaussian pump and probe beams directed at normal incidence onto a sample.  The beams are collinear and co-focused along the \emph{z}-axis.  The distance between the common beam waist and the sample surface is \emph{Z}.  The Fresnel coefficient for the reflected probe beam includes the changes due to pump-induced energy transformation processes.  The Gaussian profile of the incident pump beam is broadened by carrier diffusion in the sample.  However, the reflected AC probe beam component retains a Gaussian form with a smaller effective waist. Thus the mirror-reflected probe beam amplitude may be expanded as a sum of Gaussian beams of decreasing waist.  Given a dominant photovoltage effect according to Eq.~(\ref{eq:one}), and retaining only two terms in the expansion, the electric field of the reflected probe laser beam at the surface of the sample (disregarding the common spatial phase) may be written:
\begin{eqnarray}
E_{r} & = & \frac{E_{o}\omega_{o}}{\omega(Z)}\exp \left\{
\frac{-\rho^{2}}{\omega^{2}(Z)} \right\}  \nonumber\\
&& \times \left[\tilde{r} + \frac{\partial\tilde{r}}{\partial n}(n_{2}+ik_{2})
\frac{I_{p} \omega_{m}^{2}}{\omega_{m}^{2}(Z)} \exp \left\{
\frac{-2\rho^{2}}{\omega_{m}^{2}(Z)}\right\} \right]
\label{eq:two}
\end{eqnarray}
where $|E_{o}|^{2}$ is the intensity of the probe beam at focus, $\omega_{o}$ is the probe beam waist (\emph{i.e.} $\omega(Z) = \omega_{o}\sqrt{1+(Z/z_{o})^{2}}$, where $z_{o} = \pi\omega_{o}^{2}/\lambda$ is the Rayleigh range of the probe beam), $\rho$ is radial distance as measured from the probe beam axis, $\tilde{r}$ is the complex reflectance coefficient, $n$ is the sample refractive index, $n_{2}$ and $k_{2}$ are effective nonlinear indices defined by the coefficients appearing in Eq.~(\ref{eq:one}) (including the line-shape function), $I_{p}$ is the intensity of the pump beam at focus, and  $\omega_{m}(Z)$ is the radius of modulation as previously defined. The leading term corresponds to the DC component of the reflected beam whereas the second term corresponds to its modulated component.

Squaring the mirror-reflected probe field and integrating the over the beam profile yields the spatially integrated LPR signal via the identification:
\begin{eqnarray}
\frac{R + \Delta R}{R}=\frac{\int_{0}^{\infty} |E_{r}|^{2} \rho d\rho}{\int_{0}^{\infty} |E_{dc}|^{2} \rho d\rho}
\label{eq:three}
\end{eqnarray}
where $E_{dc}$ is just the linear reflectance amplitude. Neglecting terms second order in the nonlinear indices and performing the spatial integrations in Eq.~(\ref{eq:three}), the LPR signal may be written:
\begin{eqnarray}
\frac{\Delta R}{R}=\frac{4 n_{2} I_{p}}{n^{2}-1} \times \frac{\omega_{p}^{2}+L_{d}^{2}}{\omega^{2}(Z)+\omega_{p}^{2}(Z)+L_{d}^{2}}
\label{eq:four}
\end{eqnarray}
where $n^{2}\gg k^{2}$.  Note the \emph{Z} dependence of the LPR amplitude is contained entirely in the denominator of Eq.~(\ref{eq:four}).  The appearance of $L_{d}^2$ in the denominator shows the \emph{Z} dependence of the LPR signal will depend strongly on diffusion length.

At intermediate frequencies where the recombination lifetime $\tau$ is comparable to the modulation period, $\tau$ likewise becomes coupled into the \emph{Z} dependence of the LPR signal through the appearance of the complex diffusion length $\tilde{L}_{d}=L_{d}/\sqrt{1+i\Omega\tau}$, where $\Omega$ is the modulation frequency in radians per second. In particular, Eq.~(\ref{eq:four}) demonstrates that the LPR signal as a function of \emph{Z} may be parameterized by the expression:
\begin{eqnarray}
\frac{\Delta R}{R}=\frac{A\exp i\phi_{o}}{\omega^{2}(Z)+\omega_{p}^{2}(Z)+\tilde{L}_{d}^{2}}
\label{eq:five}
\end{eqnarray}
where $A$ and $\phi_{o}$ are the \emph{Z} independent amplitude and phase, respectively.  Thus if the pump is amplitude modulated at frequencies where $\Omega\tau \sim 1$, the recombination time will likewise become directly coupled into the \emph{Z} dependence of the phase.

If $\omega_{o}^{2}+\omega_{p}^{2} \leq |\tilde{L}_{d}|^{2}$, the 3D limit will be approached for $Z=0$. (Note Eq.~(\ref{eq:two}) is valid in the 3D limit.) However, well away from $Z=0$ (\emph{i.e.} where $\omega^{2}(Z)+\omega_{p}^{2}(Z)\geq |\tilde{L}_{d}|^{2}$), the 1D limit is restored.  The coupling of $\tilde{L_{d}}$ into the \emph{Z}-dependence of the LPR signal indicates that $L_{d}$, and ultimately $\tau$, may be determined by a regressive fit to the experimental \emph{Z}-scan LPR data.  For example, according to Eq.~(\ref{eq:five}), the \emph{Z} dependence of the LPR amplitude may be simply parameterized by the set of variables: $A$, $L_{d}^{2}$, $\omega_{p}^{2}$, $\omega_{o}^{2}$, and $\Omega\tau$, whereas the corresponding phase expression allows parametrization using the variables: $\phi_{o}$, $L_{d}^{2}$, $\omega_{p}^{2}$, $\omega_{o}^{2}$, and $\Omega\tau$.
Note that $\Omega$, $\omega_{o}$, and $\omega_{p}$ are (known) system parameters.
The parameters $A$, $L_{d}^{2}$, and $\Omega\tau$ are correlated within the amplitude fit while the parameters $\phi_{o}$ $L_{d}^{2}$, and $\Omega\tau$ are correlated in the phase fit.  However, an iterative procedure involving independent nonlinear fits to the amplitude and phase expressions may be used to establish $L_{d}$ and $\tau$ and their statistical uncertainties.  For example, the analytic expression for the phase equation may first be used to provide an independent estimate of $\tau$ and its uncertainty \cite{NumRec}.  Then the output value for $\tau$  may be held constant in the amplitude fit in order to estimate $L_{d}$ and its uncertainty.

\section{Experimental}

A set of silicon samples wafers with various \emph{p}-type ultra shallow junction structures were tested using the \emph{Z}-scan LPR technique. The shallow junctions were formed in silicon (100) substrates by implantation of \emph{n}-type dopant (As) followed by low-energy high-dose B implantation.  Dopant activation was performed using millisecond timescale flash-lamp based annealing. A range of base temperature and flash temperature targets were used to study dopant activation, dopant diffusion, and material quality.  The process conditions studied included: (i) flash target temperatures in the $1250-1350^{\circ}$C range, (ii) an additional thermal annealing of the As counter doped layer prior to B implantation, and (iii) use of a Ge amorphizing implant (AI) to reduce B ion channeling. SIMS data indicated post-activation B doping levels of $\approx 1\times 10^{19}$/cc at $X_{j}\cong 20$ nm across the sample set.  The AI process introduces a layer of crystalline defects close to the sample surface.  These defects reduce the carrier diffusion length and recombination time in the implanted region.

The \emph{Z}-scanning LPR measurement system was configured with pump and probe beam wavelengths of $488$ and $375$ nm, respectively. The pump light has an absorption depth in Si of $\approx 500$ nm. The wavelength of the probe beam is near the lowest energy direct interband transition in Si, resulting in a dominant photovoltage effect.  The optical absorption depth in Si at $375$ nm is $\cong 23$ nm.  Therefore, any detected photo-voltage must occur at or near the surface \cite{Aspnes69}.  The modulated pump ($\Omega=750$ kHz) and DC probe beams were co-focused to an $\approx2$ $\mu$m radius on the shallow junction samples.  The entire reflected probe beam was collected and focused onto the detector, thus radially integrating the beam.  The samples were scanned through focus and the \emph{Z}-scan LPR data was recorded. Estimates for $\Omega\tau$ were obtained via regressive fitting to the phase data.  Then $\Omega\tau$ was fixed in regressive fits to the amplitude data in order to yield $L_{d}^2$.  The estimated uncertainty in the extracted parameters were also output from the fitting procedure.

\begin{figure}
\includegraphics[width=246pt]{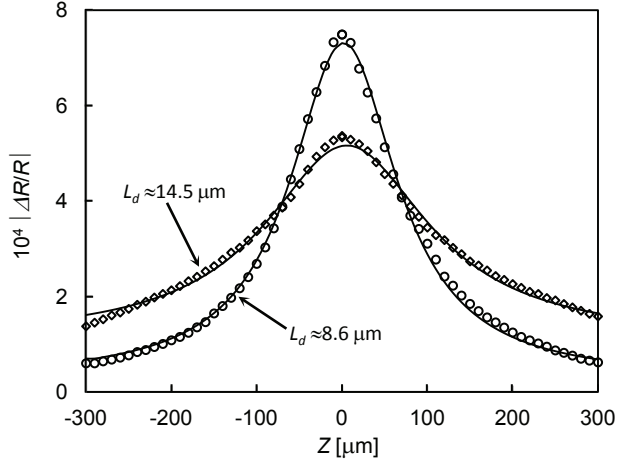}
\caption{\label{fig:Lfit}\emph{Z}-scan LPR amplitude data and fits showing the effect of near surface damage (due to AI) on shallow electrical junctions formed in silicon.  The more narrow \emph{Z} profile indicates a shorter diffusion length.}
\end{figure}

\begin{figure}[!h]
\includegraphics[width=246pt]{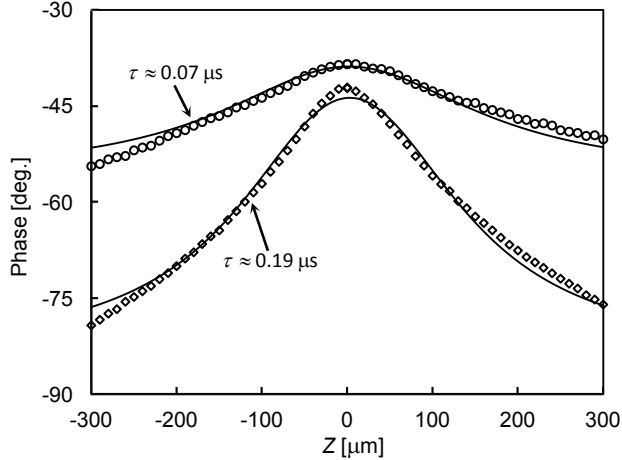}
\caption{\label{fig:taufit}\emph{Z}-scan LPR phase data from the same pair of samples as shown in Fig.~\ref{fig:Lfit}, again showing the effect of near surface damage (due to AI) on the junction.  The broader \emph{Z} profile indicates a shorter recombination lifetime.}
\end{figure}

Fig.~\ref{fig:Lfit} shows experimental Z-scan LPR amplitude data and fits obtained from samples with and without AI. The amplitudes are symmetric with respect to \emph{Z}, as expected from Eq.~(\ref{eq:five}). The more sharply peaked LPR response as a function of \emph{Z} seen on the sample with AI evidences a shorter diffusion length.  This behavior was apparent in the raw data for all samples that received the AI process, as expected. Likewise, Fig.~\ref{fig:taufit} shows experimental \emph{Z}-scan LPR phase data and fit obtained from the same pair of samples as shown in Fig.~\ref{fig:Lfit}.  The phases are again symmetric with respect to \emph{Z}, in accord with Eq.~(\ref{eq:five}).  Note the more sharply peaked amplitude data corresponds to the broader phase data. The mobility and its estimated uncertainty were obtained from the extracted parameters via the Einstein relation.  Table 1 lists fitted values of diffusion length, recombination time, and mobility for the subset of samples with AI, assuming a measurement uncertainty of $2$ ppm for the LPR amplitude and $0.13^{\circ}$ for the LPR phase.

Systematic variations in extracted parameters with process conditions are observed. The extracted carrier parameters show little sensitivity to the As thermal anneal (columns labeled ``As pre-soak'').  This is expected since the AI step occurred after the thermal As anneal (prior to B implantation).   When the $1300^{\circ}$C$/550^{\circ}$C flash anneal is repeated, the diffusion length increases by a factor of $\approx 1.5X$, while the recombination lifetimes are reduced by $\approx 10\%$, resulting in an over $2X$ increase in mobility.  When the base temperature of the flash anneal is increased to $600^{\circ}$C, the recombination time roughly doubles, indicating better removal of the AI damage.  However, the observed diffusion length only increases $\approx 10\%$.  This behavior indicates the repeated $1300^{\circ}$C$/550^{\circ}$C flash anneal achieves good junction activation but does not completely anneal the AI damage. For all samples tested, the measured mobilities agree with values expected from the activated doping levels \cite{Heilweil}. The estimated uncertainty in the extracted mobility remains less than $2\%$ in all cases.

\begin{table*}[!h]
\caption{\label{tab:table1} Measured carrier diffusion lengths, recombination times and mobilities, as determined via fitting to \emph{Z}-scan LPR data obtained from the subset of samples that received amorphizing implant.}
\begin{ruledtabular}
\begin{tabular}{ccccccc}
&\multicolumn{2}{c}{$L_{d}$ [$\mu$m]}&\multicolumn{2}{c}{$\tau$ [ns]}&\multicolumn{2}{c}{$\mu$ [$\mathrm{cm}^{2}/\mathrm{V}\cdot\mathrm{s}$]}\\
Flash temp [$^{\circ}$C]&Flash only&As pre-soak&Flash only&As pre-soak&Flash only&As pre-soak\\ \hline

$1300/550$&
$6.07\pm 0.02$& $6.01\pm 0.02$&
$90.5\pm 0.5$& $87.6\pm 0.6$&
$157\pm 2$& $159\pm 2$\\

$1300/550(2X)$&
$9.09\pm 0.03$& $8.63\pm 0.02$&
$83.1\pm 0.4$& $67.0\pm 0.5$&
$382\pm 4$ & $427\pm 5$\\

$1300/600$&
$9.36\pm 0.08$& $9.92\pm 0.08$&
$159.9\pm 0.5$& $167.3\pm 0.5$&
$211\pm 3$& $226\pm 4$\\

$1350/600$&
$10.19\pm 0.05$& $11.68\pm 0.02$&
$163.0\pm 0.5$& $167.9\pm 0.5$&
$245\pm 4$& $312\pm 4$\\

\end{tabular}
\end{ruledtabular}
\end{table*}

The \emph{Z}-scanning LPR based technique presented here has been used to characterize carrier diffusion lengths, recombination lifetimes, and mobilities with high precision.  In addition, the LPR amplitude at focus has been previously used to characterize active doping concentration (\emph{i.e.} through the dependence of Eq.~(\ref{eq:one}) on $N_{e}$) \cite{JVST}.  Therefore, provided the active dopant concentration can be determined from the LPR amplitude at focus, the mobility as measured from the \emph{Z}-scanning LPR technique may be used to characterize the sheet resistance $R_{s}$ via the relation $R_{s}\propto 1/\mu N_{e}$.


\newpage

\begin{thebibliography}{99}

\bibitem{SPV}
L. Kronik and Y. Shapira, ``Surface photovoltage phenomena:
theory, experiment, and applications,'' Surf Sci. Rep. {\bf 33}, 1 (1999).

\bibitem{SPV2}
D.K. Schroder, ``Surface voltage and surface photovoltage:
history, theory and applications,'' Meas. Sci. Technol. {\bf 12}, R16 (2001).

\bibitem{FCAetc}
H.-J. Schulze, A. Frohnmeyer, F.-J. Niedernostheide, F. Hille, P. Tutto, T. Pavelka and G. Wachutka, ``Carrier Lifetime Analysis by Photoconductance Decay and Free Carrier Absorption Measurements,'' J. Electrochem. Soc. {\bf 148}, G655 (2001).

\bibitem{PL}
D. Baek, S. Rouvimov, B. Kim, T.-C. Jo, and D.K. Schroder, ``Surface recombination velocity of silicon wafers by photoluminescence,'' Appl. Phys. Lett. {\bf 86}, 112110 (2005).

\bibitem{THz}
R. Ulbricht, E. Hendry, J. Shan, T.F. Heinz, and M. Bonn, ``Carrier dynamics in semiconductors studied with time-resolved terahertz spectroscopy,'' Rev. Mod. Phys. {\bf 83}, 543 (2011).

\bibitem{Heilweil}
B.G. Alberding, W.R. Thurber, and E.J. Heilweil, ``Direct comparison of time-resolved Terahertz spectroscopy and Hall Van der Pauw methods for measurement of carrier conductivity and mobility in bulk semiconductors,'' J. Opt. Soc. Am. B {\bf 34}, 1392 (2017).

\bibitem{Pollak}
F.H. Pollak, ``Modulation Spectrscopty of Semiconductors and Semiconductor Microstructures,'' in {\it Handbook of Semiconductors}, Vol. 2 (``Optical Properties of Semiconductors''), edited by M. Balkanski (North-Holland, Amsterdam, 1994), pp. 527-635.

\bibitem{Aspnes72}
D.E. Aspnes, ``Linearized Third-Derivative Spectroscopy with Depletion-Barrier Modulation,'' Phys. Rev. Lett. {\bf 28}, 913 (1972).

\bibitem{Aspnes69}
D.E. Aspnes and A. Frova, ``Influence of Spatially Dependent Perturbations on Modulated Reflectance and Absorption of Solids,'' Solid State Comm. {\bf 7}, 155 (1969).

\bibitem{Schroder97}
D.K. Schroder, ``Carrier Lifetimes in Silicon,'' IEEE Trans. Electron Devices {\bf 44}, 160 (1997).

\bibitem{Park}
J.E. Park, D.K. Schroder, S.E. Tan, B.D. Choi, M. Fletcher, A. Buczkowski, and F. Kirscht, ``Silicon Epitaxial Layer Lifetime Characterization,'' J. Electrochem. Soc. {\bf 148}, G411 (2001).

\bibitem{Opsal87}
J. Opsal, M.W. Taylor, W.L. Smith, and A. Rosenwaig, ``Temporal behavior of modulated optical reflectance in silicon,'' J. Appl. Phys. {\bf 61}, 240 (1987).

\bibitem{Mandelis}
A. Mandelis, {\it Diffusion-Wave Fields: Mathematical Methods and Green Functions}, (Springer, New York, 2001), pp. 626-627.

\bibitem{Opsal83}
J. Opsal, A. Rosencwaig, and D.L. Willenborg, ``Thermal-wave detection and thin-film thickness measurements with laser beam deflection,'' Appl. Opt. {\bf 22}, 3169 (1983).

\bibitem{Weaire}
D. Weaire, B. S. Wherrett, D. A. B. Miller, and S. D. Smith, ``Effect of low-power nonlinear refraction on laser-beam propagation in InSb,'' Opt. Lett. {\bf 4}, 331 (1979).

\bibitem{SBahae}
M. Sheik-Bahae, A. A. Said, and E. W. Van Stryland, ``High-sensitivity, single beam $n_{2}$ measurements,'' Opt. Lett. {\bf 14}, 955 (1989); M. Sheik-Bahae, A. A. Said, T. Wei, D. J. Hagan, and E. W. Van Stryland, ``Sensitive Measurement of Optical Nonlinearities Using a Single Beam,'' IEEE J. Quantum Electron. {\bf 26}, 760 (1990).

\bibitem{Petrov}
D.V. Petrov, A.S.L. Gomes, and C.B. de Araujo, ``Reflection \emph{Z}-scan technique for measurements of optical properties of surfaces,'' Appl. Phys. Lett. {\bf 65} 1067 (1994).

\bibitem{NumRec}
W.H. Press, S.A. Teukolsky, W.T. Vetterling, and B.P. Flannery, {\it Numerical Recipes in Fortran 77: The Art of Scientific Computing}, 2nd ed. (Cambridge University Press, New York, 1996), pp. 650-700.

\bibitem{JVST}
W. Chism, M. Current, and V. Vartanian, ``Photoreflectance characterization of ultrashallow junction activation in millisecond annealing,'' J. Vac. Sci. Technol. B {\bf 28}, C1C15 (2010).

\end{thebibliography}

\end{document}